\documentclass[pra,twocolumn,showpacs,floatfix]{revtex4}
\usepackage{graphicx}
\begin{document}

\title{Excitation spectrum of a two-component Bose-Einstein
condensate in a ring potential}
\author{J. Smyrnakis$^1$, M. Magiropoulos$^1$, A. D. Jackson$^2$,
and G. M. Kavoulakis$^1$}
\affiliation{$^1$Technological Education Institute of Crete, P.O. 
Box 1939, GR-71004, Heraklion, Greece \\
$^2$The Niels Bohr International Academy, The Niels Bohr Institute, 
Blegdamsvej 17, DK-2100, Copenhagen \O, Denmark}
\date{\today}

\begin{abstract}

A mixture of two distinguishable Bose-Einstein condensates 
confined in a ring potential has numerous interesting 
properties under rotational and solitary-wave excitation. 
The lowest-energy states for a fixed angular momentum coincide 
with a family of solitary-wave solutions. In the limit of weak 
interactions, exact diagonalization of the many-body Hamiltonian 
is possible and permits evaluation of the complete excitation 
spectrum of the system.

\end{abstract}
 \maketitle

{\it Introduction.} 
Cold atomic gases have a number of advantages that make 
them ideal for the realization and study of various 
physical effects.  Topologically nontrivial toroidal 
trapping potentials are of particular experimental and 
theoretical interest, and they have been realized 
by various experimental groups in recent years \cite{Kurn, 
Olson, Phillips1, Foot, Zoran}. 

Such trapping geometries allow the study of nonlinear
effects including the excitation of solitary waves and
of states with circulation. In a classic paper, Lieb 
considered periodic, one-dimensional motion of bosonic 
particles at zero temperature in the limit where the 
circumference of the ring, $2 \pi R$, and the number of 
particles, $N$, tend to infinity with the ratio $N/R$ 
finite and studied the excitation spectrum \cite{Lieb}. 

In the present study we consider a mixture of two 
distinguishable Bose-Einstein condensates, confined 
to a torus of a finite radius $R$, assuming 
(quasi)-one dimensional motion. In a previous study 
\cite{prl} we focused on the stability of persistent 
currents as the relative population of the two 
components is varied. Here, we demonstrate that this 
system has a rich variety of interesting properties.

Specifically, we consider solitonic and 
rotational excitations of the system. Within the 
mean-field approximation, we evaluate ``bound 
states" of solitary-wave solutions in the two 
components. A family of these solutions actually 
coincides with the states that result from minimizing 
the energy for fixed angular momentum, the ``yrast states" 
in the terminology of nuclear physics \cite{BM}. This 
identification is general and is thus valid for any 
coupling. This result generalizes that of a recent 
study for a single-component system \cite{yrsol}. 
The challenge in this identification lies in the fact that 
solitary-wave solutions propagate with a fixed 
velocity while the yrast problem is solved for fixed 
angular momentum.  As we demonstrate below, for a certain 
range of angular momentum there are ``giant" sinusoidal 
traveling-wave solutions which have a constant propagation 
velocity. We also identify various discontinuous phase 
transitions which occur in the order parameters of the two 
components as well as in the propagation velocity of the 
waves. Finally, we go beyond the mean-field approximation to 
describe the exact diagonalization of the many-body Hamiltonian 
in the limit of weak interactions and derive the full excitation 
spectrum analytically.

{\it Model.}
The problem that we have in mind is a quasi-one-dimensional
toroidal potential, with $N_A$ and $N_B$ number of atoms of 
species $A$ and $B$. Within the mean-field approximation, the 
coupled Gross-Pitaevskii equations for the order parameters 
of the two components, ${\Phi}_A$ and ${\Phi}_B$, are
\begin{eqnarray}
   i \hbar \frac {\partial \Phi_k} {\partial t} = 
- \frac {\hbar^2} {2 M R^2} \frac {\partial^2 
{\Phi}_k}{\partial \theta^2} + N U (|{\Phi}_k|^2 + 
|{\Phi}_l|^2) {\Phi}_k,
%{\mu}_k {\Phi}_k,
\label{gpe}
\end{eqnarray}
where $k=(A,B)$, $M$ is the mass of the two species 
(which is assumed the same for the two components),  
and $U = 4 \pi \hbar^2 a/(M S)$ is the matrix element 
for zero-energy elastic atom-atom collisions. (We also 
assume equal and positive scattering lengths for 
interspecies and intraspecies collisions). Further, 
$R$ is the radius of the torus, and $S$ is its cross 
section, with $R \gg \sqrt S$. Finally, $N = N_A + 
N_B$ is the total number of atoms, and the order 
parameters are normalized as $R \int |\Phi_k|^2 
d \theta = N_k/N$.

{\it Yrast states and solitary-wave solutions.}
For the yrast problem one must minimize the energy of 
the system for some fixed total angular momentum. This can  
be done with the introduction of a Lagrange multiplier. The
mathematical structure of this problem is closely related that 
for solitary-wave solutions propagating with a constant velocity.  
In fact, converting the time derivative into spatial derivative, 
the two problems are seen to be identical as we have shown 
recently in the case of a single component \cite{yrsol}.  
In the present case, the order parameters of the yrast states, 
$\Psi_A$ and $\Psi_B$, are solutions of the following coupled 
equations:
\begin{eqnarray} 
- \frac {\hbar^2} {2 M R^2} \frac {\partial^2 
{\Psi}_k}{\partial \theta^2} + N U (|{\Psi}_k|^2 + 
|{\Psi}_l|^2) {\Psi}_k - \Omega {\hat l} \Psi_k 
% \nonumber \\
 = \mu_k  \Psi_k,
\nonumber \\
\label{gpyr}
\end{eqnarray}
where $\Omega$ is the frequency of rotation of the 
density distribution of the two components, and ${\hat l}$
is the operator describing the total angular momentum of the
system (i.e., $({\hat L}_A + {\hat L}_B) \hbar $ divided by the 
total number of particles $N = N_A + N_B$).  Results identical 
to Eqs.\,(\ref{gpyr}) are obtained from Eqs.\,(\ref{gpe}) 
if one makes the solitary-wave assumption, i.e., traveling-wave 
solutions of the form $\Phi_k (\theta,t)= e^{- i \mu_k 
t/\hbar} \Psi_k(\theta - \Omega t)$. The situation is
somewhat more complicated than in the one-component case, 
since there are several families of solutions, corresponding 
to ``grey-grey", ``grey-bright", and ``bright-bright" 
excitations.  For the repulsive interactions considered here, 
the solitary-wave solutions with the lowest energy, which are 
identical to the yrast solutions, are those for which the total 
density, $n_A + n_B$, has the smallest possible variation.  
As we will show explicitly below, these are the grey-bright 
solutions.

To derive the solitary-wave solutions of the above Manakov 
system \cite{Manakov} of Eqs.\,(\ref{gpyr}), we make the 
ansatz $n_B = \kappa n_A + \lambda$, where $\kappa$ and 
$\lambda$ are parameters that can be determined from the 
consistency conditions and the constraints of the problem. 
The details of this calculation will be reported elsewhere. 
The solutions for the density are Jacobi elliptic functions, 
as is the case for a single-component Bose-Einstein condensate 
confined to a ring \cite{Carr, SMKJ}, i.e.,
\begin{eqnarray}
  n_k = n_{k,{\rm min}} + (n_{k,{\rm max}} - n_{k,{\rm min}})
  \, f_k^2 \left( \theta \frac {K({\tilde m})} {\pi} |{\tilde m} 
  \right),
\label{den}
\end{eqnarray}  
where we write $\theta$ instead of $\theta-\Omega t$ here 
and below.  In the above equation $f_A = {\rm sn}(x|{\tilde m})$ 
and $f_B = {\rm cn}(x|{\tilde m})$ are the Jacobi elliptic 
functions, and $K({\tilde m})$ is the elliptic integral 
of the first kind. The phase of the order parameters can be 
obtained from continuity equation, i.e., the equations 
$\int_{-\pi}^{\pi} \partial \varphi_k /\partial \theta =  
2 \pi  q_k$, where $\varphi_k$ is the phase of $\Psi_k$, 
$\Psi_k = \sqrt{n_k} \, e^{i \varphi_k}$, and $q_k$ is the 
winding number for each component.
 
While the Jacobi solutions are general, they imply 
complicated relationships between the various parameters. 
We now focus on several aspects of this problem which, 
although very simple, help to provide a more general picture. 
In what follows we restrict the value of the angular
momentum to the range $0 \le l \le 1$.  In a single-component 
system Bloch \cite{Bloch} has shown that the more general 
solution and the corresponding energy for any other value 
of $l$ can be evaluated from excitation of the center of 
mass motion. In the present problem with two species the 
same theorem applies \cite{prl}, and therefore one can get 
the solution for any value of $l$.

{\it An exact result.}
Assume without loss of generality that $x_B = N_B/N$ 
is smaller that $x_A = N_A/N$.  The first remarkable 
result applies when $l$ is in the range $0 \le l \le x_B$ 
and $x_A \le l \le 1$.  In this range of $l$ the exact 
solution of this problem is remarkably simple, as it is 
possible to make the total density homogeneous. The 
easiest way to derive this exact solution is to note 
that when the density $n_A(\theta)+n_B(\theta)$  is 
constant, the coupled, nonlinear Gross-Pitaevskii 
equations, Eqs.\,(\ref{gpe}), become linear equations. 
(We stress that identical expressions also follow from 
the solitary-wave solutions of Eqs.\,(\ref{den}) with 
$\kappa$ of our ansatz equal to $-1$). More specifically, 
since the total density is constant, the solution of 
Eqs.\,(\ref{gpyr}) is 
\begin{eqnarray}
  \Psi_A = \sqrt{x_A} \, (c_0 \phi_0 + c_m \phi_m),
  \Psi_B = \sqrt{x_B} \, (d_0 \phi_0 + d_m \phi_m),
\label{yrst2}
\end{eqnarray}
where $\phi_m = e^{i m \theta}/\sqrt{2 \pi R}$. The 
corresponding coefficients are given by
\begin{eqnarray}
  c_0 = \sqrt{\frac {(x_A - l/m) (1 - l/m)} 
  {x_A (1 - 2 l/m)}}, \ 
  c_m = \sqrt{\frac {(x_B - l/m) l/m} 
  {x_A (1 - 2 l/m)}},
\nonumber \\
\end{eqnarray}
Similar results for the $B$ component are obtained by 
interchanging the indices $A$ and $B$ interchanged and 
with $d_m$ negative. Here, $l$ is in the interval $0 \le 
l \le m$. Each component has a sinusoidal density distribution,
\begin{eqnarray}
   n_k(\theta) &=& N_k (1 + 2 {\cal C}_k \cos m \theta)/
   (2 \pi R),
\label{dens1} 
\end{eqnarray}
with ${\cal C}_A = c_0 c_m$ and ${\cal C}_B = d_0 d_m$.
Since $x_A c_0 c_m + x_B d_0 d_m = 0$, the total density 
$n_A + n_B$ is indeed constant and equal to $N/(2 \pi R)$. 
Also
\begin{eqnarray}
  \tan \varphi_k = \frac {\sin m \theta} 
{{\cal D}_k + \cos m \theta}, 
\label{phase2}
\end{eqnarray}
where ${\cal D}_A = c_0/c_m$ and ${\cal D}_B = d_0/d_m$.
The energy per particle is given by $E/(\epsilon N) 
= \gamma/2 + m l$, where $\gamma/2 = 2 N a R/S$ is the 
ratio of the interaction energy of the cloud 
with a homogeneous density $N/(2 \pi R)$ of $N = N_A 
+ N_B$ atoms and the kinetic energy $\epsilon = 
\hbar^2/(2 M R^2)$.  However, since $m$ is an integer, 
the solution with the lowest energy, or equivalently 
the solution with the lowest velocity of propagation 
$u = \Omega R$, is the one with $m=1$.  Thus, 
$u = c \equiv \hbar/(2 M R)$, which is constant, 
simply because the dispersion relation scales linearly 
with the angular momentum $l$, i.e., $E/(\epsilon N) = 
\gamma/2 + l$. While the above solution is clearly the 
yrast state, the phase-space constraints do not allow 
it to have an arbitrary value of $l$ but rather restrict 
it to the range $0 \le l \le x_B$ and $x_A \le l \le 1$. 
From Eqs.\,(\ref{phase2}) it follows that (for $m=1$) the 
winding number of the smaller component $B$ changes at the 
values of the angular momentum $ l_1 = (1 - \sqrt{1 - 2 
x_B})/2$ and $l_2 = (1 + \sqrt{1 - 2 x_B})/2$. As we see 
below, it also changes at the symmetry point $l = 1/2$.  
(Obviously, it has to change in the interval $x_B \le l \le 
x_A$). 

\begin{figure}[t]
\includegraphics[width=8.cm,height=5cm]{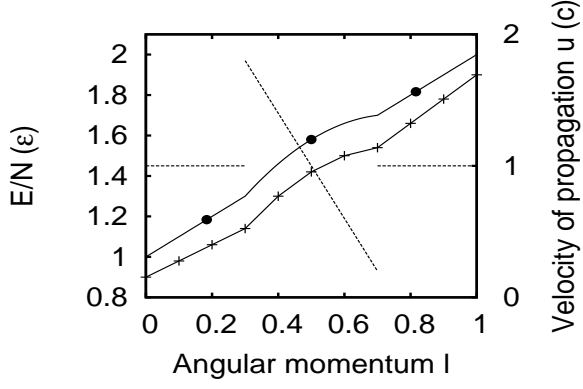}
\caption{The energy of the gas within the mean-field 
approximation (solid curve) and the lowest eigenvalues
of the Hamiltonian of Eq.\,(\ref{ham}) (crosses). The
dashed curve gives the derivative of the mean-field 
energy, i.e., the velocity of propagation of the waves.
Here $N_A = 7$ ($x_A = 0.7$), $N_B = 3$ ($x_B = 0.3$), 
$N = 10$, and $\gamma = 2$.   (Although this is a 
perturbative result and $\gamma \ll 1$, we have chosen 
this value in order to make the curves more visible.)  
The three dots are the points with $l=l_1, 1/2$, and 
$l_2$.}
\label{FIG1}
\end{figure}

{\it Full solution for weak interactions.}
In addition to the exact solution that we have found in the 
range $0 \le l \le x_B$ and $x_A \le l \le 1$, the problem
also has a simple solution in the range $x_B \le l \le 
x_A$ in the limit of weak coupling.  Again, the solitary-wave 
and yrast solutions coincide. From the point of view 
of Eqs.\,(\ref{den}), the parameter $\tilde{m}$ tends to zero 
in the limit of weak interactions, and one can expand 
the solution in powers of $\tilde{m}$.  The Jacobi
functions become sinusoidal since only the single-particle 
orbitals $\phi_0$ and $\phi_1$ contribute to the order 
parameters. In this case the parameters of our ansatz 
are $\kappa = -x_B/x_A$ and $\lambda = x_B/(\pi R)$. 
The presence of $\phi_0$ and $\phi_1$ only is reasonable 
if one thinks in terms of the yrast states, where minimization of 
the energy in the limit of weak interactions is accomplished with 
these two single-particle states of lowest kinetic energy.  
(Inclusion of $\phi_{-1}$ is also energetically expensive since 
it does not have the same sign as the angular momentum). The 
order parameters are of the same form as Eqs.\,(\ref{yrst2}), 
with
\begin{eqnarray}
   c_0 = d_1 = \sqrt{\frac {x_A - l} {x_A - x_B}}, \,\,\,
   c_1 = - d_0 = \sqrt{\frac {-x_B + l} {x_A - x_B}}.
\end{eqnarray}
The corresponding density and the phase of the two components 
is given by Eqs.\,(\ref{dens1}) and (\ref{phase2}).
Remarkably, in the specific case $l=1/2$, we find that 
$\Psi_A = {\sqrt{x_A}}(\phi_0 + \phi_1)/{\sqrt {2}}$,
and $\Psi_B = {\sqrt{x_B}}(-\phi_0 + \phi_1)/{\sqrt {2}}$.
The point $l=1/2$ is singular in the sense that for $l \to 
(1/2)^{-}$, the winding numbers $(q_A, q_B)$ are $(0,1)$ 
and for $l \to (1/2)^{+}$, they are $(1,0)$. The winding 
numbers thus take the following values:
\begin{eqnarray}
0 \le l < (1 - \sqrt{1 - 2 x_B})/2: (q_A,q_B) = (0,0).
\nonumber \\
(1 - \sqrt{1 - 2 x_B})/2 < l < 1/2: (q_A,q_B) = (0,1).
\nonumber \\
1/2 < l < (1 + \sqrt{1 - 2 x_B})/2: (q_A,q_B) = (1,0).
\nonumber \\
(1 + \sqrt{1 - 2 x_B})/2 < l \le 1: (q_A,q_B) = (1,1).
\end{eqnarray}
The energy is quadratic in $l$ in this case,
\begin{eqnarray}
 E /({\epsilon N}) - {\gamma}/{2} = l 
+ {\gamma} (x_A - l) (l - x_B). 
\end{eqnarray}
Its slope, or equivalently the velocity $u$ of propagation, 
varies linearly with $l$, $u/c = 1 + {\gamma} 
(1 - 2 l)$.  At $l= x_A$ it changes discontinuously by 
an amount ${|\delta u|}/c = {\gamma} (x_A - x_B)$.

While the results presented above for the range $x_B 
\le l \le x_A$ are only valid for weak interactions, the more
general yrast/solitary-wave solutions are given in terms 
of the Jacobi elliptic functions, as described above. 
For stronger couplings, the picture presented for the 
winding numbers remains the same. The density distribution 
may be either sinusoidal or exponentially localized, 
depending on the value of the parameters, which 
determine the value of $\tilde{m}$.  Also as noted 
above, for $\tilde{m} \to 0$ the Jacobi functions 
become sinusoidal, and for $\tilde{m} \to 1$ they become 
exponentially localized.

{\it Exact solution of the many-body Hamiltonian for
weak interactions.}

What has been presented so far is based on the mean-field
approximation, which assumes that the many-body state is 
a product state. As we saw earlier, in the limit of weak 
interactions one may work with the single-particle states 
$\phi_0$ and $\phi_1$ only.  In this case the 
many-body Hamiltonian can be diagonalized exactly.  We start 
with the Hamiltonian in second quantized form, where 
$\hat{c}_m, \hat{c}_m^{\dagger}, \hat{d}_m$, and 
$\hat{d}_m^{\dagger}$ with $m=0,1$ are annihilation and 
creation operators of the species $A$ and $B$, respectively,
\begin{eqnarray}
  \hat{H} = 
  \epsilon (\hat{c}_1^{\dagger} \hat{c}_1 + 
  \hat{d}_1^{\dagger} \hat{d}_1) +
%\nonumber \\ + 
\frac v {2} (\hat{c}_0^{\dagger} \hat{c}_0^{\dagger} 
\hat{c}_0 \hat{c}_0 
+ \hat{d}_0^{\dagger} \hat{d}_0^{\dagger} \hat{d}_0 \hat{d}_0 
\nonumber \\
+ \hat{c}_1^{\dagger} \hat{c}_1^{\dagger} \hat{c}_1 \hat{c}_1 
+ \hat{d}_1^{\dagger} \hat{d}_1^{\dagger} \hat{d}_1 \hat{d}_1
+ 4 \hat{c}_0^{\dagger} \hat{c}_0 \hat{c}_1^{\dagger} \hat{c}_1
+ 4 \hat{d}_0^{\dagger} \hat{d}_0 \hat{d}_1^{\dagger} \hat{d}_1
\nonumber \\
+ 2 \hat{c}_0^{\dagger} \hat{c}_0 \hat{d}_0^{\dagger} \hat{d}_0 
+ 2 \hat{c}_0^{\dagger} \hat{c}_0 \hat{d}_1^{\dagger} \hat{d}_1
+ 2 \hat{c}_1^{\dagger} \hat{c}_1 \hat{d}_0^{\dagger} \hat{d}_0 
+ 2 \hat{c}_1^{\dagger} \hat{c}_1 \hat{d}_1^{\dagger} \hat{d}_1
\nonumber \\
+ 2 \hat{c}_0 \hat{c}_1^{\dagger} \hat{d}_0^{\dagger} \hat{d}_1 
+ 2 \hat{c}_0^{\dagger} \hat{c}_1 \hat{d}_0 \hat{d}_1^{\dagger}), 
\end{eqnarray}
where $v= U/(2 \pi R S)$.  If one considers the algebra of the 
bilinears of the annihilation-creation operators that appear in 
the Hamiltonian, one can recognize two copies of the SU(2) 
algebra and two central elements, corresponding to the particle 
numbers of the two species. Let us introduce the operators 
$\hat{n}_{A,0} = \hat{c}_0^{\dagger} \hat{c}_0$, $\hat{n}_{A,1} 
= \hat{c}_1^{\dagger} \hat{c}_1$, $\hat{n}_{B,0} = 
\hat{d}_0^{\dagger} \hat{d}_0$, $\hat{n}_{B,1} = \hat{d}_1^{\dagger} 
\hat{d}_1$, and $\hat{j}_A = \hat{c}_1^{\dagger} \hat{c}_0,\ 
\hat{j}_B = \hat{d}_1^{\dagger} \hat{d}_0$.  The two central 
elements are $\hat{n}_{A,0}+ \hat{n}_{A,1}=N_A$ and $\hat{n}_{B,0}
+\hat{n}_{B,1}=N_B$.  If one defines $\hat{j}_{A,0} = (\hat{n}_{A,1} 
- \hat{n}_{A,0})/2$, $\hat{j}_{B,0} = (\hat{n}_{B,1} - \hat{n}_{B,0})/2$, 
the two copies of the SU(2) algebra are generated by the operators 
$\{\hat{j}_A, \hat{j}_A^{\dagger}, \hat{j}_{A,0}\}$, $\{ \hat{j}_B, 
\hat{j}_B^{\dagger}, \hat{j}_{B,0}\}$, where $[\hat{j}_{A,B,0},
\hat{j}_{A,B}]=\hat{j}_{A,B}$ and $[\hat{j}_{A,B},
\hat{j}_{A,B}^{\dagger}]=2\hat{j}_{A,B,0}$. The total spin operators 
(Casimir elements) for the two algebras are given by 
$\hat{\underline{j}}_{A,B}^2 = \hat{j}_{A,B,0}^2 + 
(\hat{j}_{A,B}\hat{j}_{A,B}^{\dagger} + \hat{j}_{A,B}^{\dagger}
\hat{j}_{A,B})/2$.

With the above definitions, the operator that measures the angular 
momentum in units of $\hbar$ takes the form $\hat{L} = N/2 + 
\hat{j}_{A,0}+ \hat{j}_{B,0}$, which allows us to write the 
Hamiltonian in the form
\begin{eqnarray}
  \hat{H} = \epsilon {\hat{L}}  
  &+& v \Big{[} \frac{1}{2}N \left( N - 1 \right)-\frac{1}{2}N_AN_B 
  - {\hat{L}^2} + N {\hat{L}} \nonumber \\
 &+& \left((\hat{\underline{j}}_{A}+
 \hat{\underline{j}}_{B})^2-\hat{\underline{j}}_{A}^2-
 \hat{\underline{j}}_{B}^2\right) \Big{]} .
\label{ham0}
\end{eqnarray}

Since the possible eigenvalues of $\hat{j}_{A,0}$, $\hat{j}_{B,0}$ 
run from $-N_{A,B}/2$ to $N_{A,B}/2$, we are in the spin $j_{A,B}=
N_{A,B}/2$ representation of SU(2). Hence $\hat{\underline{j}}_{A,B}^2
=N_{A,B}(N_{A,B}/2+1)/2$. Also, the spin of the 
$\hat{\underline{j}}_{A}+ \hat{\underline{j}}_{B}$ representation 
satisfies $(N_A-N_B)/2 \le j_{AB} \le N/2$. This means that the 
eigenvalues of the Hamiltonian for a state of definite angular 
momentum $L$ assume the form
\begin{eqnarray}
  E = \epsilon {L}  
  + v \left( N \left( \frac N 4 - 1 \right) 
  + j_{AB} (j_{AB} + 1) 
  - {L^2} + N {L} \right). 
\nonumber \\
\label{ham}
\end{eqnarray}
The minimum value of the energy is attained for the minimum value 
of $j_{AB}$ that is compatible with the given value of the angular 
momentum $L$. One must distinguish three cases. For $0 \le L \le N_B$, 
then $j_{AB} = N/2 - L + k$.  As a result the $k$th excited state 
(with $k = 0,1,2,\dots L$) is 
\begin{eqnarray}
E_k - E_0 = (\epsilon - v) L + v k (N - 2L + k + 1).
\end{eqnarray}
For $N_B \le L \le N_A$, then $j_{AB} = (N_A - N_B)/2 + k$, 
and thus
\begin{eqnarray}
E_k - E_0 = \epsilon L + v [(N_A - L) (L - N_B) 
- N_B] 
\nonumber \\
+ v k (N_A - N_B + k + 1). 
\label{mid}
\end{eqnarray}
For $N_A \le L \le N$, we find that $j_{AB} = L - N/2 + k$, which
implies that
\begin{eqnarray}
E_k - E_0 = (\epsilon + v) L + v [- N 
+ k (2 L - N + k + 1)],
\end{eqnarray}
where $E_0 = v N (N - 1)/2$ is the energy of the
nonrotating cloud. The above expressions agree with those 
of the mean-field approximation given above to leading order 
in $N$  for $k=0$. Figure 1 shows the lowest eigenvalues for 
$N_A = 7$, $N_B = 3$ for the choice $\gamma = 2$. Clearly, the 
slope of the linear parts has a correction which is of order 
$1/N$.

The case of equal populations $N_A = N_B = N/2$ with an 
angular momentum $N/2$ is easy to describe and rather
peculiar.  Within the mean-field description, all the angular 
momentum is transferred from one component to the other 
as $l$ crosses the value $1/2$.  Beyond mean-field, it is 
convenient to work in the representation of the Fock states
$| m \rangle = |0^{N/2 - m}, 1^{m} \rangle \bigotimes
|0^{m}, 1^{N/2 - m} \rangle$. Here the first ket refers 
to component $A$ and the second to $B$. In this notation 
$N/2 - m$ atoms of species $A$ reside in the state $\phi_0$, 
etc. It is readily shown that the state
\begin{eqnarray}
 | \Psi \rangle =  \frac 1 {\sqrt{N/2 + 1}} \sum_{m=0}^{N/2}
 (-1)^m | m \rangle,
\end{eqnarray}
is an eigenstate, and actually it is the yrast state, with
an eigenvalue $\epsilon N/2 + N (N-2) v/2$, in agreement with
Eq.\,(\ref{mid}). This is a highly correlated state and all 
the four eigenvalues of the density matrix are equal to $N/4$.

{\it Conclusions and overview.} 
A mixture of two distinguishable Bose-Einstein condensed 
gases which are confined in a ring potential shows a 
remarkable collection of interesting effects. Although 
this system is relatively complicated, we have succeeded 
in obtaining a number of simple analytic results. In 
addition we have shown that the yrast states coincide 
with a certain class of solitary-wave solutions. 

For a range of angular momentum the two species develop 
a quite unusual structure: Although they have a long-wavelength 
and sinusoidal density distribution, extending over the whole 
range of the ring (independent of the value of the coupling), 
they have a finite amplitude and in that respect differ from 
the sound waves we are familiar with. In addition, their 
velocity of propagation is constant, and in the limit of a 
large ring, this velocity tends to zero. When phase-space 
constraints do not allow the angular momentum to take the 
required value, the total density can no longer remain 
constant and the velocity of propagation changes 
discontinuously. The rather complicated solutions in 
this range of the angular momentum may be expressed in terms 
of Jacobi elliptic functions. In the limit of weak coupling 
they take again a very simple form. In the same limit we have
provided the full spectrum of the many-body Hamiltonian. 

The two-component system that we have considered is in no 
sense a trivial extension of the single-component system. The 
combination of two degrees of freedom associated with the two 
species, with a topologically-nontrivial motion make this problem 
very interesting. The presence of a second component gives the 
system the freedom to make its total density as homogeneous as 
possible and thus to reduce its interaction energy. This is what 
lies behind many of the peculiar properties described above. The 
problem examined here is in principle more general and thus more 
complicated than its single-component counterpart. In spite of 
this rich structure, it nevertheless has a less complicated 
solution.  We believe that experimental confirmation of the 
effects predicted would be both a challenging and a rewarding 
task.

This project is implemented through the Operational Program 
"Education and Lifelong Learning", Action Archimedes III and 
is co-financed by the European Union (European Social Fund) 
and Greek national funds (National Strategic Reference Framework 
2007 - 2013). We also acknowledge the ``POLATOM" ESF Research 
Networking Programme for its support.

\end{document}